\documentclass[a4paper,twocolumn,twoside,fleqn]{article}
\usepackage[T1,T2A]{fontenc}
\usepackage[cp1251]{inputenc}
\usepackage[UKenglish,german,ukrainian,russian]{babel}
\usepackage[colorlinks,urlcolor=blue,pagebackref]{hyperref}
\usepackage{cite}
\usepackage{balance}
\usepackage{fancyhdr}
\usepackage{amsmath}
\usepackage{xy}
\xyoption{matrix}\xyoption{arrow}
\renewenvironment{abstract}{%
\footnotesize
}{\par}
\newdimen\labelindent\labelindent=\parindent\advance\labelindent by \labelwidth\advance\labelindent by \labelsep
\newcounter{listcounter}
\newcommand{\softm}[1]{#1{\kern-1.3pt}'{\kern-.3pt}}
\newcommand{\softL}{L{\kern-2pt}'}
\newcommand{\ms}{\kern.3em}
\newcommand{\nbs}{\kern5pt}
\newcommand{\RN}[1]{\uppercase\expandafter{\romannumeral#1}}

\newtheorem{prop}{Proposition}
\newtheorem{define}{Definition}
\newcommand{\NO}[1]{N$^{\rm o}$\kern.1em #1}
\font\BBB=msbm10    
\font\ssb=cmssbx10 
\font\sssm=cmss8 
\newcommand{\A}{\mbox{\ssb A}}
\newcommand{\B}{\mbox{\ssb B}}
\newcommand{\CC}{\mbox{\ssb c}}
\newcommand{\V}{\mbox{\ssb v}}
\newcommand{\K}{\mbox{\ssb k}}
\newcommand{\Zero}{\mbox{\ssb 0}}
\newcommand{\eisf}[1]{\mbox{\sssm#1}}
\newcommand{\bcdot}{\mbox{\boldmath$\cdot$}}
\newcommand{\bpartial}{\mbox{\boldmath$\partial$}}
\newcommand{\bmat}[1]{\mbox{\boldmath$#1$}}
\newcommand{\ubm}[1]{\mbox{\unboldmath$#1$}}
\newcommand{\bkey}[1]{\hbox{\boldmath$#1$}}
\newbox\abox\setbox\abox=\hbox{\ssb A}
\newbox\Elocity\setbox\Elocity=\hbox{\ssb v}
\newbox\PrimE\setbox\PrimE=\hbox{{\ssb v}$^{\prime}$}
\newbox\elocity\setbox\elocity=\hbox{\sssm v}
\newbox\Prime\setbox\Prime=\hbox{{\sssm v}$^{\scriptscriptstyle\prime}$}
\catcode`A=13\def A{\copy\abox}\catcode65=11
\catcode`V=13\def V{\copy\Elocity}\catcode86=11
\catcode`W=13\def W{\copy\PrimE}\catcode87=11
\catcode`v=13\def v{\copy\elocity}\catcode118=11
\catcode`w=13\def w{\copy\Prime}\catcode119=11
\catcode`D=13\let D=\wedge\catcode68=11
\catcode`M=13\def M{\bf\Omega}\catcode77=11
\newcommand{\bcode}{\begingroup\catcode`A=13 \catcode`V=13
\catcode`W=13 \catcode`M=13 \catcode`v=13 \catcode`w=13\catcode`D=13\catcode`"=13}
\newcommand{\e}{\endgroup}
\newdimen\colsep\colsep=\arraycolsep
\newcommand{\SSS}[1]{{\scriptscriptstyle#1}}
\newcommand{\au}{\|\bu\|}
\newcommand{\bfe}{\hbox{\ssb e}}
\newcommand{\x}{{\sf x}}
\newcommand{\sv}{{\sf v}}
\newcommand{\sfp}{{\sf p}}

\newcommand{\sE}{{\sf E}}
\newcommand{\se}{{\sf e}}
\newcommand{\sA}{{\sf A}}

\newcommand{\sk}{{\sf k}}
\newcommand{\bu}{\mbox{\boldmath$u$}}
\newcommand{\bvarepsilon}{\mbox{\boldmath$\varepsilon$}}
\newcommand{\bepsilon}{\mbox{\boldmath$\epsilon$}}
\newcommand{\R}{\mbox{\BBB R}}
\newcommand{\BBN}{\mbox{\BBB N}}

\newcommand{\sR}{\boldsymbol{\mathsf r}}
\newcommand{\sP}{\boldsymbol{\mathsf p}}
\newcommand{\sV}{\boldsymbol{\mathsf v}}
\newcommand{\sX}{\boldsymbol{\mathsf x}}
\newcommand{\tsx}{\tilde{\mathsf x}}
\newcommand{\tsv}{\tilde{\mathsf v}}
\renewcommand{\sp}{\mathsf p}
\newcommand{\sr}{\mathsf r}
\newcommand{\tsR}{\tilde{\boldsymbol{\mathsf r}}}
\newcommand{\tsP}{\tilde{\boldsymbol{\mathsf p}}}
\newcommand{\tsp}{\tilde{\mathsf p}}
\newcommand{\tsr}{\tilde{\mathsf r}}
\newcommand{\tH}{\tilde{H}}
\newcommand{\tpi}{\tilde{\pi}}
\newcommand{\trho}{\tilde{\rho}}
\newcommand{\tbQ}{\tilde{\boldsymbol{Q}}}
\newcommand{\bP}{{\boldsymbol{P}}}
\newcommand{\bQ}{{\boldsymbol{Q}}}
\newcommand{\tbP}{\tilde{\boldsymbol{P}}}
\newcommand{\tQ}{\tilde{Q}}
\newcommand{\tP}{\tilde{P}}
\begin{document}
\fancyfoot[LE,RO]{\selectlanguage{UKenglish}\bf\thepage}
\fancyfoot[LO,RE]{{\selectlanguage{ukrainian}\small\it ISSN 0503-1265 Укр. фіз. журн. Т.~48, №~4}}
\selectlanguage{UKenglish}
\author{\href{http://iapmm.lviv.ua/12/eng/files/st_files/matsyuk.htm}{R.\ms Ya.\ms Matsyuk}}
\title{\vspace*{-16mm}\leftline{\scriptsize UDC~[531.12/.13.01:530.12]:[531.314.2:512.816.7]:517.972.7}Differential geometric
mechanisms in Ostrohrad\softm skyj relativistic spherical top dynamics\vspace*{-3mm}}
\date{\vspace*{-3mm}\href{http://www.iapmm.lviv.ua/12/eng/index.htm}{Institute for Applied Problems in Mechanics \& Mathematics}
    \\15 Dudayev St., 79005 \softL viv, Ukraine
    \\matsyuk@lms.lviv.ua\\
    \url{http://www.iapmm.lviv.ua}
   \vspace*{-5mm}
    }
\maketitle
\thispagestyle{fancy}
\begin{abstract}
Some intrinsic tools from the formal theory of variational equations are being demonstrated at work in application to one concrete example of the third-order evolution equation of free relativistic top in three-dimensional space-time. The main goal is to introduce a combined approach consisting in the simultaneous utilization of symmetry principles along with the inverse variational problem considerations in terms of vector-valued differential forms.
Next, some simple algorithm of transition between the autonomous variational problem and the variational problem in parametric form is established. The example definitely solved shows  no-existence of a globally and intrinsically defined Lagrangian for the Poincar\'e-invariant and well defined unique variational equation in the case in hand. Hamiltonian counterpart is briefly discussed in terms of Poisson bracket. The model appears to provide a generalized canonical description of the quasi-classical spinning particle governed by the Mathisson-Papapetrou equations in flat space-time.
\end{abstract}
\section*{Introduction}
Ostrohrad\softm skyj's mechanics has been repeatedly
revisited from the point of view of global analysis
including certain features of intrinsic differential geometry
(see monographs \cite{Krup,Leon,Saund}, preceded and followed by large number
of other reviews and articles). The differential geometry of Os\-tro\-hrad\-\softm s\-kyj's mechanics
gained a renewed interest on the part of contemporary mathematicians during
 past two decades.

On the other hand, applications of the higher-order variational calculus to some classical
models of a relativistic particle motion began in 1937 and continue
till now. The investigations on the application of Os\-t\-ro\-hrad\softm s\-kyj's mechanics to
real physical models haven't been abandoned since the pioneer works by
Mathisson, Bopp, Weyssenhoff, Raabe, and H\"onl (see \cite{Math,Bopp,Weyss,Hoenl}. Most of the applications consider models of test particles
endowed with inner degrees of freedom~\cite{Tulcz1,Riewe,Plyush,Nest,Leiko,Arodz,Ners,Arreaga} or models
which put the notion of the acceleration onto the framework of general
differential geometric structure of the extended configuration space of
the particle~\cite{Scarp}. One interesting example of how the derivatives
of the third order appear in the equations of motion of test particle is
provided by Mathisson-Papapetrou equations
\begin{eqnarray}
  &&  \frac{D}{d\zeta}\left(m_{\SSS0}\frac{u^\alpha}{\au}+\frac{u_\gamma
}{\au^2}\frac{D}{d\zeta}S^{\alpha\gamma}\right) = {\cal F}^\alpha
\label{MPu}
\\ && \frac{D}{d\zeta}S^{\alpha\beta} = \frac{1}{\au^2}\left(u^\beta
u_\gamma\frac{D}{d\zeta}S^{\alpha\gamma}-u^\alpha
u_\gamma\frac{D}{d\zeta}S^{\beta \gamma}\right) \label{MPS}
\end{eqnarray}
together with the supplementary condition
\begin{equation}
u_\gamma S^{\alpha\gamma}=0.
\label{Pir}
\end{equation}

It is immediately clear that the second term in (\ref{MPu}) may produce
the derivatives of the third order of space-time variables $x^\alpha$ as
soon as one dares to substitute
$\displaystyle u_\gamma\frac{\strut DS^{\alpha\gamma}}{d\zeta}$ by $\displaystyle -
S^{\alpha\gamma}\frac{\strut Du^\gamma}{d\zeta}$ in virtue of (\ref{Pir}). Such
substitution in fact means differentiating equation (\ref{Pir}). However,
the system of equations thus obtained will not possess any additional
solutions comparing to that of (\ref{MPu}--\ref{Pir}) as far as one does
not forget of the original constraint (\ref{Pir}). The system (\ref{MPu}--\ref{Pir}) was recently a subject of discussion in~\cite{PlyaPhRev}.
In (\ref{MPu}) the right hand side vanishes if there is no gravitation.

It is a matter of common consent that the relativistic motion of simple
particles in gravitational field may be described mathematically via the
notion of geodesic paths. Because less simple particles obey higher-order
equations of motion, it seems worthwhile to investigate
the appropriate geometries. But, in the same way as pseudo-Riemannian
geometry descends down to the natural representation of Lorentz group,
more complicated geometry should break out first from some symmetry
considerations of global character.

We intend to present in this contribution some tools from the arsenal of
intrinsic analysis on manifolds that may appear helpful in solving the
invariant inverse problem of the calculus of variations. In special case
of three-dimensional space-time we shall successfully follow some
prescriptions for obtaining a third-order Poincar\'e-invariant variational
equations up to the very final solution thus discovering the unique possible one, which
will then be identified with the motion of free relativistic top by means
of comparing it to (\ref{MPu}--\ref{Pir}) when
$R^\alpha{}_{\beta\delta\gamma}=0$. This case of two-dimensional
motion in space makes quite a good sense from the viewpoint of the general
theory as well~\cite{PlyaBook}. On the other hand, one can show directly that
even in four-dimensional special relativity case the world line of a
particle obeying the system of equations (\ref{MPu}--\ref{Pir}) has the
third curvature equal to zero (see also~\cite{Yakup}). Thus, even in this
case the
particle actually propagates in two-dimensional space. Another feature of this
limited case is that the spin four-vector
\begin{equation}
\sigma_{\alpha}=\frac{1}{2\au}\epsilon_{\alpha\beta\gamma\delta}u^\beta
S^{\gamma\delta}
\label{s}
\end{equation}
keeps constant under the condition of the motion
be free. So knowing a Lagrange function for some third-order equation,
equivalent to (\ref{MPu}--\ref{Pir}), allows offering a generalized
Hamiltonian description in terms of Poisson bracket that might be
considered as a canonical equivalent to (\ref{MPu},~\ref{Pir}).
Our example exposes some typical features of variational calculus:
\begin{itemize}
\item the nonexistence (in our case) of well defined invariant
Lagrangian along with intrinsically very well defined equation of motion
with Poincar\'e symmetry produced by each of a family of degenerate Lagrangians
which transform into each other by renumbering the axes of Lorentz frame;
\item all handled Lagrangians give rise to the same system of canonical equations;
\item each  Lagrangian includes different set of se\-cond order
derivatives, thus their sum is not a Lagrangian of minimal order.
\end{itemize}

\section{Homogeneous form and pa\-ra\-me\-t\-ric invariance}
Presentation of the equation of motion in so-called `manifestly
covariant form' stipulates introducing of the space of
Ehresmann's velocities of the configuration manifold $M$ of the particle,
$T^kM=\{x^\alpha,\dot x^\alpha,\ddot x^\alpha\dots x_{(k)}^\alpha\}$. In
future the notations $u^\alpha$, $\dot u^\alpha$, $\ddot u^\alpha$,
$u_{(r)}^\alpha$ will frequently be used in place of $\dot x^\alpha$,
$\ddot x^\alpha$, $x_{(3)}^\alpha$, $x_{(r+\SSS1)}^\alpha$, and also
$x_{(\SSS0)}^\alpha$ sometimes will merely denote the $x^\alpha$. We call
some mapping $\zeta\mapsto x^\alpha(\zeta)$ the {\it parametrized\/}
(by means of
$\zeta$) {\it world line\/} and its image in $M$ will be called
the {\it non-parameterized world
line.} As far as we are interested in a variational equation (of order
$s$) that would describe the non-parameterized world lines of the
particle,
\begin{equation}
{\cal E}_\alpha\left(
x^\alpha,u^\alpha,\dot u^\alpha,\ddot
u^\alpha,\dots, u_{(s-\SSS1)}^\alpha
\right)=0,
\label{f}
\end{equation}
the Lagrange function ${\cal L}$  has to satisfy the Zermelo conditions,
which in our case of only up to the second order derivatives present in
${\cal L}$ read
\begin{eqnarray*}
   &   & u^\beta\frac{\partial}{\partial u^\beta}{\cal L}+2\dot
u^\beta\frac{\partial}{\partial \dot u^\beta}{\cal L}={\cal L}
\\ &   & u^\beta\frac{\partial}{\partial \dot u^\beta}{\cal L}=0.
\label{}
\end{eqnarray*}

In this approach the independent variable $\zeta$ (called the {\it parameter along
the world line}\/) is not included into the configuration manifold $M$. Thus
the space $T^kM$ is the appropriate candidate for the role of the
underlying manifold on where the variational problem in the autonomous
form should be posed. We may include the parameter $\zeta$ into the
configuration manifold by introducing the trivial fibre manifold
$\R\times M\to\R$, $\zeta\in\R$, and putting into consideration its
$k^{\rm th}$-order prolongation, $J^k(\R,M)$, i.~e. the space,
constituted by the $k^{\rm th}$-order jets of local cross-sections of
$Y=\R\times M$ over $\R$. Each such cross-section of $Y$ is nothing but
the graph in $\R\times M$ of some local curve $x^\alpha(\zeta)$ in $M$.
For each $r\in\BBN$ there exists an obvious projection
\begin{equation}
p_{\SSS0}^r:J^r(\R,M)\to T^rM
\label{p0r}
\end{equation}
as follows. The manifold $T^rM$ consists of the derivatives up to the
$r^{\rm th}$-order of curves $x^\alpha(\zeta)$ in $M$, evaluated at
$0\in\R$. If for every $\tau\in\R$ we denote by same character $\tau$
the mapping $\zeta\mapsto\zeta+\tau$ of
$\R$ onto itself, then the projection reads
\begin{eqnarray}
\nonumber &&p_{\SSS0}^r:\biggl(\tau;x^\alpha(\tau),\frac{d}{d\zeta}x^\alpha(\tau),
\frac{d^2}{d\zeta^2}x^\alpha(\tau),\dots
\\&&\hspace{4cm}\dots,\frac{d^r}{d\zeta^r}x^\alpha(\tau)\biggr)\mapsto{}\nonumber
\\&& {}\mapsto
\nonumber\biggl(\,(x^\alpha\circ\tau)\,(0), \frac{d}{d\zeta}\,(x^\alpha\circ\tau)\,(0),
\\\label{p0} &&\hspace{1cm}\frac{d^2}{d\zeta^2}\,(x^\alpha\circ\tau)\,(0), \dots,
 \frac{d^r}{d\zeta^r}\,(x^\alpha\circ\tau)\,(0)\biggr).
\end{eqnarray}

By means of the projection (\ref{p0r},~\ref{p0}) every Lagrange function
${\cal L}$ initially defined on $T^kM$ may be pulled back to the manifold
$J^k(\R,M)$ and defines there some function ${\cal L}_{{\SSS0}}$ by the obvious
formula ${\cal L}_{{\SSS0}}=\nobreak{\cal L}\circ p_{\SSS0}^k$. We say that the
differential form
\begin{equation}
\lambda={\cal L}_{\SSS0}d\zeta
\label{lambda0}
\end{equation}
constitutes a variational problem in extended parametric form because in
the construction of the new configuration manifold $\R\times M$ the
independent variable $\zeta$ was artificially doubled. But we shall need
this construction later.

Let us return to the variational problem set on the manifold $T^kM$ by
a given Lagrange function ${\cal L}$. The very first moment we impose the
Zermelo conditions, the problem becomes degenerate. There exists one way
to avoid degeneracy by reducing the number of velocities. Of course, at
the cost of losing the ``homogeneity'' property of the equation~(\ref{f}). Consider
some way of segregating the variables $x^\alpha\in M$ into $t\in\R$
and $\x^i\in Q$, $\dim Q=\dim M-1$, thus making $M$ into some fibration,
$M\approx\R\times Q$, over $\R$. The manifold of jets $J^r(\R,Q)$
provides some local representation of what is known as the manifold
$C^r(M,1)$ of $r$-contact one-dimensional submanifolds of $M$.
Intrinsically defined global projection of non-zero elements of $T^rM$
onto the manifold $C^r(M,1)$ in this local and, surely, ``non-
covariant'' representation is given by
\begin{equation}
\wp^r:T^rM\setminus\{0\}\to J^r(\R,Q),
\label{wp}
\end{equation}
and in the third order is implicitly defined by the following formulae,
where the local coordinates in  $J^r(\R,Q)$ are denoted by $t; \x^i,
\sv^i, {\sv'}^i, {\sv''}^i, \dots, \sv_{(r-\SSS1)}^i$ with $\sv_{(\SSS0)}^i$
marking $\sv^i$ sometimes,
\begin{eqnarray}
&&\dot t\, \sv^i    =  u^i  \nonumber
\\ &&(\dot t)^3 {\sv'}^i  =   \dot t \dot u^i - \ddot t u^i \label{calp}
\\ &&(\dot t)^5 {\sv''}^i  =  (\dot t)^2 \ddot u^i - 3 \dot t \ddot t \dot
u^i + \left[3(\ddot t)^2 -\dot t t_{(\SSS3)}\right] u^i.   \nonumber
\end{eqnarray}

There does not exist any well-defined projection from the manifold
$C^r(M,1)$ onto the space of independent variable $\R$, so the expression
\begin{equation}
\Lambda=L\left(t; \x^i, \sv^i, {\sv'}^i, {\sv''}^i, \dots, \sv_{(k-
\SSS1)}^i \right)dt
\label{L}
\end{equation}
will vary in the dependence on the concrete way of local representation $M\approx\R\times
Q$. We say that two different expressions of type~(\ref{L}) define one
and the same variational problem in parametric form if their difference
expands into nothing but only the pull-backs to $C^k(M,1)$ of the following
contact forms, which live on the manifold $C^{\SSS1}(M,1)$,
\begin{equation}
\theta^i=d\x^i-\sv^idt.
\label{Theta}
\end{equation}
These differential forms obviously vanish along the jet of every curve
$\R\to Q$.

Let the components of the variational equation
\begin{equation}
\sE_i = 0
\label{fe}
\end{equation}
of the Lagrangian (\ref{L}) be treated as the components of the following vector one-
form,
\begin{equation}
\bfe=\big\{\sE_idt\big\}.
\label{e}
\end{equation}

We intend to give a ``homogeneous'' description to (\ref{e}) and
(\ref{L}) in terms of some objects that would live on $T^sM$ and $T^kM$
respectively. But we cannot apply directly the pull-back operation to the
Lagrangian~(\ref{L}) because the pull-back of one-form is a one-form
again, and what we need on $T^kM$ is a Lagrange {\em function}, not a
differential form. However, it is possible to pull (\ref{L}) all the way
back along the composition of projections (\ref{p0r}) and (\ref{wp}),
\begin{equation}
pr^k=\wp^k\circ p_{\SSS0}^k,
\label{p}
\end{equation}
ultimately to the manifold $J^k(\R,M)$. In such a way we obtain the
differential form $\big(L\circ pr^k\big)dt$. But what we do desire, is a
form that should involve $d\zeta$ solely (i.~e. semi-basic with respect
to the projection $J^k(\R,M)\to\R$). Fortunately, the two differential
forms, $dt$ and $\dot t d\zeta$, differ not more than only by the
contact form
\begin{equation}
\vartheta=dt-\dot t d\zeta.
\label{theta}
\end{equation}
Now, we recall that equivalent Lagrangians
that have the structure of  (\ref{L}) differ by multiplies of the contact
forms (\ref{Theta}). It remains to notice that, by the course of
(\ref{p0}) and (\ref{calp}), the pull-backs of the
contact forms (\ref{Theta}) expand only into contact forms (\ref{theta})
and
\begin{equation}
\vartheta^i=dx^i-u^i d\zeta
\label{thetas}
\end{equation}
alone,

\[
pr^{\SSS1*}\theta^i= dx^i - \big(\sv^i\circ pr^{\SSS1}\big)dt = \vartheta^i -
\big(\sv^i\circ pr^{\SSS1}\big)\vartheta.
\]
Thus, every variational problem, posed on $J^k(\R,Q)$ and represented by
(\ref{L}), transforms into an equivalent variational problem
\begin{equation}
\lambda=\big(L\circ pr^k\big)\dot t d\zeta,
\label{lambda}
\end{equation}
posed on $J^k(\R,M)$. But the Lagrange function of this new variational
problem,
\begin{equation}
{\cal L}_{\SSS0}=\big(L\circ pr^k\big)\dot t,
\label{L0}
\end{equation}
does not depend upon the parameter $\zeta$ and substantially may be
thought of as a function, defined on $T^kM$.

We prefer to cast the variational equation (of some order $s\leq 2k$),
generated by the Lagrangian (\ref{lambda}), into the framework of
vector-valued exterior differential systems theory by introducing the
following vector differential one-form, defined on the manifold $J^s(\R,M)$,
\begin{equation}
\bvarepsilon={\cal E}_\alpha \left(x^\alpha, \dot x^\alpha, \dots,
x_{(s)}^\alpha \right)d\zeta.
\label{epsilon}
\end{equation}
The expressions ${\cal E}_\alpha \left(x^\alpha, \dot x^\alpha, \dots,
x_{(s)}^\alpha \right)$ in (\ref{epsilon}) may also be treated as ones,
defined on $T^sM$, similar to ${\cal L}_{\SSS0}$. Altogether the
constructions, built above, allow formulation of the following statement:
\begin{prop}\label{homogen}
If the differential form (\ref{e}) corresponds to the variational
equation of the Lagrangian (\ref{L}), then the expressions
\begin{equation}
{\cal E}_\alpha = \left\{-u^i\sE_i, \dot t \,\sE_i \right\}
\label{CalE}
\end{equation}
correspond to the Lagrange function (\ref{L0}).
\end{prop}

In this case the ($s^{\rm th}$-order) equation (\ref{f}) describes ``in
homogeneous form'' the same non-parameterized world lines of a particle
governed by the variational problem
(\ref{L0}), as does the equation (\ref{fe}) with the Lagrangian given by
(\ref{L}), and also ${\cal L}_{\SSS0}$ obviously satisfies the Zermelo
conditions. As to more sophisticated details, paper~\cite{DGA2001} may be consulted.

\section{Criterion of variationality}
Our main intention is to find a Poincar\'e-invariant ordinary (co-vector)
differential equation of the third order in three-dimensional space-time.
With this goal in mind we organize the expressions $\sE_i$ in (\ref{e})
into a single differential object, the exterior one-form
\begin{equation}
\se_{\SSS0}=\sE_i d\x^i
\label{e0}
\end{equation}
defined on the manifold $J^s(\R,Q)$, so that the vector differential form
(\ref{e}) should now be treated as the coordinate representation of the
intrinsic differential geometric object
\begin{equation}
\se=\se_i d\x^i=\sE_i dt \otimes d\x^i = dt \otimes \se_{\SSS0}.
\label{efull}
\end{equation}
This way constructed differential form $\se$ is an element of the graded
module of differential semi-basic with respect to $\R$ differential forms
on $J^s(\R,Q)$ with values in the bundle of graded algebras $\wedge T^*Q$ of scalar forms on
$TQ$. Of course, due to the dimension of $\R$, actually only functions
(i.~e. semi-basic zero-forms) and semi-basic one-forms (i.~e. in $dt$ solely) exist. We also wish to mention
that every (scalar) differential form on $Q$ is naturally treated as a
differential form on $T^rQ$, i.~e. as an element of the graded algebra of cross-sections of
$\wedge T^*\big(T^rQ\big)$.

For arbitrary $s\in\BBN$ let $\Omega_s(Q)$ denote the algebra of (scalar) differential forms on
$T^sQ$ with coefficients depending, aside of $\sv_{r-\SSS1}$,  $r\le s$, also on $t\in \R$. It is
possible to develop some calculus in $\Omega_s(Q)$ by introducing the
operator of vertical (with respect to $\R$) differential $d_v$
and the operator of total (or formal ``time'') derivative $D_t$ by means
of the prescriptions:

\[
d_v f =\frac{\partial f}{\partial \x^i} d\x^i + \frac{\partial
f}{\partial \sv_{(r)}^i} d\sv_{(r)}^i, \quad {d_v}^2 = 0,
\]
so that $d_v\x^i$ and $d_v\sv_{(r)}^i$ coincide with $d\x^i$ and
$d\sv_{(r)}^i$ respectively, and

\[
D_tf=\frac{\partial f}{\partial t}+ \sv^i \frac{\partial f}{\partial
\x^i}+ \sv_{(r+\SSS1)}^i \frac{\partial f}{\partial \sv_{(r)}^i}, \quad
D_td_v = d_vD_t.
\]

There exists a notion of {\it derivation} in graded algebras endowed with
generalized commutation rule, as  $\Omega_s(Q)$ is. An operator $D$ is
called a derivation of degree $q$ if for any differential form $\varpi$
of degree $p$ and any other differential form $w$ it is true that
$D(\varpi \wedge w) = D(\varpi) \wedge w +(-1)^{pq} \varpi \wedge D(w)$.
To complete the above definitions it is necessary to demand that $d_v$ be
a derivation of degree $1$ whereas $D_t$ be a derivation of degree $0$.
But still this is not the whole story. We need one more derivation of
degree $0$, denoted here as $\iota$, and defined by its action on
functions and one-forms, which altogether locally generate the algebra
$\Omega_s(Q)$,

\[
\iota f = 0, \iota d\x^i = 0, \,\iota d\sv^i = d\x^i,
\iota d\sv_{(r)}^i = (r+1)\, d\sv_{(r-\SSS1)}^i.
\]

Let operator $\deg$ mean evaluating of the degree of a differential form.
The {\em Lagrange differential} $\delta$ is first introduced by its
action upon the elements of $\Omega_s(Q)$,

\[
\delta = \left(\deg + \sum_{r=1}^s\frac{(-1)^r}{r!}D_t{}^r \iota^r \right)d_v,
\]
and next trivially extended to the whole of the graded module of semi-basic
differential forms on $J^s(\R,Q)$ with values in  $\wedge
T^*\big(T^rQ\big)$ by means of
\begin{eqnarray*}
&&\delta (\omega_i dt\otimes d\x^i)  =  dt\otimes \delta (\omega_i d\x^i),
 \\ &&\delta (\omega^r_i dt\otimes d\sv_{(r)}^i)  =  dt\otimes \delta
(\omega^r_i d\sv_{(r)}^i).
\end{eqnarray*}

This $\delta$ turns out to possess the property $\delta^2=0$.  We have,
that for the differential geometric objects (\ref{efull}) and (\ref{L})
the following relation holds:
\begin{equation}
\se=\delta\Lambda=dt\otimes\delta L.
\label{edelta}
\end{equation}
Now the criterion for an arbitrary set of expressions $\big\{\sE_i\big\}$
in (\ref{e})  be the variational equations for some Lagrangian reads
\begin{equation}
\delta \se = dt\otimes\delta \se_{\SSS0} =0,
\label{crit}
\end{equation}
with $\se$ constructed from $\big\{\sE_i\big\}$ by means of (\ref{e0})
and (\ref{efull}).

Of course, one may apply the above constructions literally to the
analogous objects living on the manifold $J^s(\R,M)$ in (\ref{p0r}) and
obtain the operator, the Lagrange differential, $\delta^{\SSS Y}$ acting
upon semi-basic, with respect to $\R$, differential forms on  $J^s(\R,M)$
with values in the bundle $\wedge T^*\big(T^sM\big)$. In the algebra
$\Omega_s(M)$ operator $\delta^{\SSS Y}$ preserves the sub-algebra of forms
that do not depend on the parameter $\zeta\in\R$. The restriction of
$\delta^{\SSS Y}$ to the algebra of differential forms truly defined on
$T^sM$ sole will be denoted by $\delta^{\SSS T}$. It was introduced
in~\cite{Tulcz2}. If in (\ref{lambda0}) the Lagrange function ${\cal L}_{\SSS0}$
does not depend on the parameter $\zeta\in\R$, as is the case of
(\ref{lambda}--\ref{L0}), rather than applying $\delta^{\SSS Y}$ to the forms $\lambda$
from (\ref{lambda0}) and
\begin{equation}
\varepsilon = \varepsilon_\alpha dx^\alpha ={\cal E}_\alpha d\zeta\otimes
dx^\alpha
\label{epsfull}
\end{equation}
from (\ref{epsilon}), we may apply the restricted operator  $\delta^{\SSS
T}$ to the Lagrange function ${\cal L}_{\SSS0}$ and to the differential
form
\begin{equation}
\varepsilon_{\SSS0} = {\cal E}_\alpha dx^\alpha.
\label{eps0}
\end{equation}
In case of (\ref{L0}) the criteria $\delta^{\SSS Y}\varepsilon=0$,
\begin{equation}
\delta^{\SSS T}\varepsilon_{\SSS0}=0,
\label{criteps}
\end{equation}
and (\ref{crit}) are all equivalent, and the variational equations,
produced by the expressions $\varepsilon=\delta^{\SSS Y}\lambda$ from
(\ref{epsfull},~\ref{lambda}), $\varepsilon_{\SSS0}=\delta^{\SSS T}{\cal
L}_{\SSS0}$ from (\ref{eps0},~\ref{L0}), and $\se$ from (\ref{edelta}) all are
equivalent to (\ref{f}). The expressions (\ref{e}) and (\ref{L}) are not
``generally covariant'' whereas (\ref{eps0}) is. But the criterion
(\ref{criteps}) needs to be solved along with Zermelo conditions, whereas
(\ref{crit}) is self-contained.

The presentation of a system of variational expressions
$\big\{\sE_i\big\}$ under the guise of a semi-basic (i.~e. in
$dt$ solely) differential form that takes values in the bundle of one-forms
over the configuration manifold $Q$ is quite natural:
\begin{itemize}
\item the Lagrange density (called {\it Lagrangian} in this work) is a
one-form in $dt$ only;
\item the destination of the Euler-Lagrange expressions in fact consists
in evaluating them on the infinitesimal variations, i.~e. the vector
fields tangent to the configuration manifold $Q$  along the critical
curve; consequently, the set of $\sE_i$ constitutes a linear form on the
cross-sections of  $TQ$ with the coefficients depending on higher
derivatives
\end{itemize}

More details can be found in ~\cite{Kolar} and ~\cite{Var}.

\section{Lepagean equivalent}
The system of partial differential
equations, imposed on $\sE_i$, that arises from (\ref{crit}) takes more
tangible shape in the concrete case of third-order Euler-Poisson (i.~e. ordinary
Euler-Lagrange) expressions. The reader may
consult~\cite{mathmeth} and references therein. Let skew-symmetric matrix
$\A$, symmetric matrix $\B$, and a column $\CC$ all depend on $t$, $\x^i$,
and $\sv^i$ and satisfy the following system of partial differential
equations,
{\renewcommand{\arraystretch}{1.7}
\begin{equation}\label{5}
\begin{array}{l}
        \partial_{_{_{_{{\hbox{\sssm v}}}}}}{\!}_{[i}{}{{\sf A}}_{jl]}=0\,,
\\         2\,{{\sf B}}_{[ij]}-3\,{\bf D_{_{\bmat 1}}}{\kern.01667em}{{\sf
A}}_{ij}=0\,,
\\      2\,\partial_{_{_{_{{\hbox{\sssm v}}}}}}{\!}_{[i}{}{{\sf B}}_{j]l}
               -4\,\partial_{_{_{_{{\hbox{\sssm x}}}}}}{\!}_{[i}{}{{\sf
A}}_{j]l}
               +{\partial_{_{_{_{{\hbox{\sssm x}}}}}}{\!}_{l}}{\,}{{\sf A}}_{ij}+{}
\\ \hspace{4cm}
             {} +2\,{\bf D_{_{\bmat
1}}}{\kern.01667em}{\partial_{_{_{_{{\hbox{\sssm v}}}}}}{\!}_{l}}{\,}{{\sf
A}}_{ij}=0\,,
\\        {\partial_{_{_{_{{\hbox{\sssm v}}}}}}{\!}_{(i}}{}{{\sf c}}_{j)}
               -{\bf D_{_{\bmat 1}}}{\kern.01667em}{{\sf B}}_{(ij)}=0\,,
\\        2\,{\partial_{_{_{_{{\hbox{\sssm
v}}}}}}{\!}_{l}}{\,}\partial_{_{_{_{{\hbox{\sssm v}}}}}}{\!}_{[i}{}{{\sf
c}}_{j]}
           -4\,\partial_{_{_{_{{\hbox{\sssm x}}}}}}{\!}_{[i}{}{{\sf B}}_{j]l}
           +{{\bf D_{_{\bmat 1}}}}^{2}{\,}{\partial_{_{_{_{{\hbox{\sssm
v}}}}}}{\!}_{l}}{\,}{{\sf A}}_{ij}+{}
\\  \hspace{4cm}  {} +6\,{\bf D_{_{\bmat 1}}}{\kern.0334em}\partial_{_{_{_{{\hbox{\sssm
x}}}}}}{\!}_{[i}{}{{\sf A}}_{jl]}=0\,,
\\       4\,\partial_{_{_{_{{\hbox{\sssm x}}}}}}{\!}_{[i}{}{{\sf c}}_{j]}
           -2\,{\bf D_{_{\bmat 1}}}{\kern.0334em}\partial_{_{_{_{{\hbox{\sssm
v}}}}}}{\!}_{[i}{}{{\sf c}}_{j]}
           -{{\bf D_{_{\bmat 1}}}}^{3}{\,}{{\sf A}}_{ij}=0\,,
\end{array}
\end{equation}}
\noindent where the differential operator $\bf D_{_{\bmat 1}}$ is the lowest order
generator of the
Cartan distribution,
\[
{\bf D_{_{\bmat 1}}}=\partial_{t}{\,+\,}\V{\,\bkey.\,}{\bpartial}_{\eisf x}\,.
\]

It is obvious and commonly well known that the Euler-Lagrange expressions
are of affine type in the highest derivatives.
The most general form of the Euler-Poisson equation of the third order reads:
\begin{equation}\label{4}
\A{\,\bkey.\,}\V^{{{\prime}}{{\prime}}}{\, +
\,}(\V^{{\prime}}{\!\bkey.\,}{\bpartial}_{\eisf v})\,
\A{\,\bkey.\,}\V^{{\prime}}{\, + \,}\B{\,\bkey.\,}\V^{{\prime}}{\, +
\,}\CC\, = \,\Zero\,.
\end{equation}

Due to the affine structure of the left hand side of equation (\ref{4}),
we may alongside with the differential form (\ref{efull}) introduce next
one, the coefficients of which do not depend on third-order derivatives,
\begin{eqnarray}
\epsilon=\sA_{ij} d\sv^{\prime}{}^j\otimes d\x^i& {}+{}& \sk_i dt \otimes
d\x^i\,,\nonumber
\\ &&\hspace{-1.3em}\bcode\K = (W{\!\bkey.\,}\bpartial_{v})\,A{\,\bkey.\,}W+\B{\,\bkey.\,}W+\CC\e\,.
\label{6}
\end{eqnarray}

From the point of view of searching only holonomic local curves in
$J^3(\R,Q)$ those exterior differential systems who differ not more than
merely by multipliers of the contact forms (\ref{Theta}) and

\[
\theta'{}^i=d\sv^i-\sv'{}^idt, \quad \theta''{}^i=d\sv'{}^i-\sv''{}^idt,
\]
are considered equivalent. The differential forms (\ref{6}) and
(\ref{efull}) are equivalent:

\[
\epsilon-\se= \sA_{ij}\theta''{}^j\otimes d\x^i.
\]
The differential form (\ref{6}) may be accepted as an alternative
representation of the {\it Lepagean equivalent}~\cite{Krup} of
(\ref{efull}).
\section{Invariant\nbs Euler-Poisson equation}
We are preferably interested in those variational equations that expose some
symmetry.
Let $X(\bepsilon)$ denote the component-wise action of an infinitesimal
generator $X$ on a vector differential form $\bepsilon$. That the exterior
differential system, generated by the form $\bepsilon$, possesses the
symmetry of $X$ means that there exist some matrices ${\bf\Phi}$,
${\bf\Xi}$, and ${\bf\Pi}$ depending on $\copy\Elocity$ and
$\copy\PrimE$, such that
\boldmath
\begin{equation}\label{7}
\bcode \ubm X(\epsilon)={\bf\Phi}\,.\,\epsilon + {\bf\Xi}\,.\,(\hbox{\ssb x}-
V\ubm d\ubm t)
+ {\bf\Pi}\,.\,(\ubm d V - W\ubm d \ubm t)\e\ubm.
\end{equation}
\unboldmath

Equation (\ref{7}) expresses the condition that two vector exterior
differential systems, the one, generated by the vector differential form
$\bepsilon$, and the other, generated by the shifted form $X(\bepsilon)$,
are algebraically equivalent. For systems, generated by one-forms (as in
our case) this is completely the same thing as to demand that the set of
local solutions be preserved under the one-parametric Lie subgroup
generated by $X$. We see two advantages of this method:
\begin{itemize}
\item the symmetry conception is formulated in reasonably most general form;
\item the problem of invariance of a differential equation is reformulated
in algebraic terms by means of undetermined coefficients ${\bf\Phi}$,
${\bf\Xi}$, and ${\bf\Pi}$;
\item the order of the underlying non-linear manifold is reduced
(to $J^2(\R,Q)$ instead of $J^3(\R,Q)$).
\end{itemize}

Further details may be found in~\cite{Sym}.

In the case of the Poincar\'e group we assert that $\A$ and $\K$ in
(\ref{6}) do not depend upon $t$ and $\hbox{\ssb x}$. And for the sake of
reference it is worthwhile to put down the general expression of the generator
of the Lorentz group, parameterized by a skew-symmetric matrix
${\bf\Omega}$ and some vector
\boldmath
$\pi$:
\bcode
\begin{eqnarray*}
\ubm X &=&
{}-(\pi\cdot\hbox{\ssb
x})\,\ubm{\partial_t}+\ubm{g_{\SSS0\SSS0}\,t}\,\pi\,.\,\partial_{\eisf x}
+M\cdot(\hbox{\ssb x}D\partial_{\eisf x})+{}
\\ &&{}+\ubm{g_{\SSS0\SSS0}}\,\pi\,.\,\partial_{v}
+(\pi\cdot V)\:V\,.\,\partial_{v}+M\cdot(VD\partial_{v})+{}
\\ &&{}+\ubm2\,(\pi\cdot V)\:W.\,\partial{_{w}}+(\pi\cdot
W)\:V\,.\,\partial{_{w}}+{}
\\&& {}+M\cdot(WD\partial{_{w}})\,\ubm.
\end{eqnarray*}
\e
\unboldmath
Here the centred dot symbol denotes the inner product of vectors or
tensors and the lowered dot symbol denotes the contraction of a row-vector
and the subsequent column-vector.

System of equations (\ref{5},~\ref{7}) may possess many solutions. Or no solutions
at all, depending on the dimension of the configuration manifold. For
example, in dimension one the skew-symmetric matrix $\A$ does not exist.
If $\dim Q=3$, there is no solution to the P.D.E. system
(\ref{5},~\ref{7}) (see~\cite{thesis}). Fortunately, if $\dim Q=2$, the
solution exists {\em and is unique}, up to a single scalar parameter
$\mu$ (see also~\cite{Condenced} ):
\begin{prop}
The invariant Euler-Poisson 
equation 
of the relativistic two-dimensional
motion is:
\bcode
\begin{eqnarray}
\lefteqn{ -\frac{\astV''\strut}{(1+V\bcdotV)^{3/2}\strut} +
3\,\frac{\astW\strut}{(1 + V\bcdotV)^{5/2}}\,(V\bcdotW) - {}}
\nonumber
\\
&   &{} -
\frac{\mu\strut}{(1+V\bcdotV)^{3/2}}\,\bigl((1+V\bcdotV)\,W-
(W\bcdotV)\,V\bigr)=\Zero\,.
\label{10}
\end{eqnarray}
\e
\end{prop}

The dual vector above is defined in commonly used notations, $(\ast\mbox{\ssb
w})_{i}=\epsilon_{ji}{\sf w}^{j}$. We know two different Lagrange
functions for the left hand side of (\ref{10}):
\begin{eqnarray}
   L_1& =  & -\frac{\sv'^2\sv^1}{\sqrt{1+\sv_i\sv^i}(1+\sv_2\sv^2)} + \mu
\sqrt{1+\sv_i\sv^i}  \label{L1}
\\ L_2& =  &  \phantom{-} \frac{\sv'^1\sv^2}{\sqrt{1 + \sv_i\sv^i}(1 +
\sv_1\sv^1)} + \mu \sqrt{1 + \sv_i\sv^i}  \label{L2}.
\end{eqnarray}
These should differ by a total time derivative
\begin{equation}\label{matsyuk:F}
L_2-L_1=\frac{d}{dt}F\,.
\end{equation}
In fact, let $g_{ij}={\rm diag}\big( 1, \eta_1, \eta_2\big)$, $\eta_i=\pm1$. Then, if, for example, $\eta_1 \eta_2 =1$, then
\begin{equation}\label{matsyuk:ethaofonesign}
    F=\arctan\frac{\sv^1\sv^2}{\sqrt{1 + \sv_i\sv^i}}.
\end{equation}
With the help of the prescriptions of Proposition~\ref{homogen} we immediately
obtain the
`homogeneous' counterpart of~(\ref{10}):
\boldmath
\begin{eqnarray}
\lefteqn{-\frac{{\bf\ddot{\bmat u}}\times u\strut}{\|u\|^{\ubm{\scriptstyle3}}\strut}
+ 3\,\frac{{\bf\dot{\bmat u}} \times u\strut}{\|u\|^{\ubm{\scriptstyle5}}}\,({\bf\dot{\bmat u}}\cdot u) - {}} \nonumber
\\ & &{} - \frac{\ubm\mu\strut}{\|u\|^{\ubm{\scriptstyle3}}}\,\bigl( (u\cdot u)\,{\bf\dot{\bmat u}}
- ({\bf\dot{\bmat u}} \cdot u)\,u  \bigr) = 0\, \label{homvareq}
\end{eqnarray}
\unboldmath
with the corresponding family of Lagrange functions,
\begin{eqnarray*}
   {\cal L}_1 & = & \frac{u^1\big(\dot u^0 u^2-\dot u^2 u^0\big)}{\au
\big(u^0 u^0 +u_2 u^2 \big)} + \mu\au\,,
\\ {\cal L}_2 & = & \frac{u^2\big(\dot u^1 u^0-\dot u^0 u^1\big)}{\au
\big(u_0 u^0 + u_1 u^1 \big)} + \mu\au\,,
\\ {\cal L}_0 & = & \frac{u^0\big(\dot u^2 u^1-\dot u^1 u^2\big)}{\au
\big(u_1 u^1 + u_2 u^2 \big)} + \mu\au\,,
\end{eqnarray*}
where, for the sake of the `coordinate homogeneity', the notation $u^0$ was introduced to substitute the evolution of the time coordinate $\dot t$. One obtains the third expression for ${\cal L}_0$ by simple abuse of cyclic symmetry philosophy.

The difference between ${\cal L}_2$ and ${\cal L}_1$ is readily obtained from  the Proposition~\ref{homogen} again. Thus in the case when~(\ref{matsyuk:ethaofonesign}) holds, one gets from (\ref{calp}) and (\ref{p})
\begin{eqnarray*}
{\cal L}_2-{\cal L}_1 &\,=\,&u^0\bigl(L_2\circ pr^2-L_1\circ pr^2\bigr) \\
 &\,=\,&\frac{d}{d\zeta}\arctan\frac{\bigl(\sv^1\circ pr^2\bigr)\bigl(\sv^2\circ pr^2\bigr)}{\sqrt{1+\bigl(\sv_i\circ pr^2\bigr)\bigl(\sv^i\circ pr^2\bigr)}} \\
 &\,=\,& \frac{d}{d\zeta}\arctan\frac{u^1u^2}{u_0\sqrt{u_\alpha u^\alpha}}\,,
\end{eqnarray*}
and for two other differences by direct calculation and the trigonometric identity for $\arctan$:
\begin{eqnarray*}
    {\cal L}_1-{\cal L}_0 &\,=\,& \frac{d}{d\zeta}\arctan\frac{u^0u^1}{u_2\sqrt{u_\alpha u^\alpha}}\,,\\
    {\cal L}_0-{\cal L}_2 &\,=\,& \frac{d}{d\zeta}\arctan\frac{u^0u^2}{u_1\sqrt{u_\alpha u^\alpha}}\,.
\end{eqnarray*}

To produce a variational equation of the third order, the Lagrange function
should be of affine type in second derivatives. It makes no sense to even try
finding a Poincar\'e-invariant such Lagrange function in space-time dimensions
greater than two~\cite{thesis}. But the generalized momentum
\boldmath
\[
\frac{\ubm{\partial\cal L}}{\ubm\partial u} - \frac{\ubm d}{\ubm{
d\zeta}}\frac{\ubm{\partial\cal L}}{\ubm\partial {\bf\dot{\bmat u}}}
= \frac{{\bf\dot{\bmat u}}\times u}{\|u\|^{\ubm{\scriptstyle3}}} + \ubm\mu\,\frac{u}{\|u\|}
\]
\unboldmath
does not depend on the particular choice of one of the above family of Lagrange
functions. This expression for the generalized momentum was (in different
notations) in fact obtained in~\cite{Plyush} by means of introducing an
abundance of Lagrange multipliers into the formulation of the corresponding
variational problem.

\subsection{Free relativistic top in two dimensions}
This equation (\ref{homvareq}) carries certain amount of physical sense. We
leave it to the reader to ensure (see also~\cite{Astro}) that in terms of spin
vector (\ref{s}) the Mathisson-Papapetrou equations (\ref{MPu}--\ref{MPS}) under
the Mathisson-Pirani auxiliary condition (\ref{Pir}) are equivalent to the next
system of equations,
\begin{eqnarray}
&&{\varepsilon_{\alpha\beta\gamma\delta}\ddot u^\beta u^\gamma
\sigma^\delta -
3\,\frac{ \bmat{ {\bf\dot{\bmat u}} \!\cdot\! u} }{\phantom{^2}\au^2}
\,\varepsilon_{\alpha\beta\gamma\delta}\dot u^\beta u^\gamma \sigma^\delta
+ {}}\nonumber
\\&&\qquad{}+\frac{m_{\SSS0}}{\sqrt{|g|}} \left[\bmat{({\bf\dot{\bmat u}}
\cdot u)}\,u_\alpha - \au^2 \dot u_\alpha\right]  =  {\cal F}_\alpha
\label{18}
\\&& \au^2 \dot\sigma_\alpha + \bmat{(\sigma\cdot {\bf\dot{\bmat u}})}\, u_\alpha  =   0
\nonumber
\\&& \bmat{\sigma\cdot u}   =    \bmat0\,.\nonumber
\end{eqnarray}
It should be clear that the four-vector $\bmat\sigma$ is constant in all its
components if the force $\cal F_\alpha$ vanishes. Equation (\ref{18}) admits a
planar motion, when $u_3=\dot u_3=\ddot u_3=0$, and, if we put $g_{\alpha\beta}={\rm diag}\big( 1, \eta_1, \eta_2, \eta_3
\big)$, it takes
the shape of
\boldmath
\begin{eqnarray*}
&&\ubm{\eta_3\sigma_3}\left(\frac{{\bf\ddot{\bmat u}}\times u\strut}{\|u\|^{\ubm{\scriptstyle3}}} -
3\,\frac{{\bf\dot{\bmat u}}
\times u\strut}{\|u\|^{\ubm{\scriptstyle5}}}\,({\bf\dot{\bmat u}}\cdot u)\right) + {}
\nonumber
\\ &&\hfill+ {}  \frac{\ubm{m_{\SSS0}}\strut}{\|u\|^{\ubm{\scriptstyle3}}}\,\left[ (u\cdot u)\,{\bf\dot{\bmat u}} -
({\bf\dot{\bmat u}}\cdot u)\,u  \right] = 0\,,
\end{eqnarray*}
\unboldmath
where the vector $\bmat u$ becomes three-dimensional.
Comparing with (\ref{homvareq}) imposes
$\displaystyle\mu=\frac{\strut m_{\SSS0}}{\eta_3\sigma_3}$.

\section{Poisson structure}
In constructing Hamilton equations we follow the prescriptions of~\cite{Gitman}.
First, let us introduce, along with the variables $\x^i$, $\sv^i$, ${\sv'}^i$, yet more variables, $\sp_i$ and $\sr_i$. We define the following Hamilton function $H$ on the total set of variables $\x^i, \sv^i, {\sv'}^i, \sp_i$ and $\sr_i$,
\begin{equation}\label{matsyuk:HamFunction}
H=\sP.\sV+\sR.\sV'-L\,.
\end{equation}
The system of higher-order equations of motion in these variables take the shape
\begin{alignat}{2}\label{holonomic}
  \frac{d\sX}{dt} &=\phantom{-} \frac{\partial H}{\partial \sP}\,, &\quad\frac{d\sV}{dt}&=\frac{\partial H}{\partial \sR}\,,\\
  \label{Euler-Poisson}
  \frac{d\sP}{dt} &= -\frac{\partial H}{\partial \sX}\,,&& \\
  \label{OstrohradDefinition}
  \frac{d\sR}{dt} &= -\frac{\partial H}{\partial \sV}\,,  &\quad\frac{\partial H}{\partial \sV'} &= \boldsymbol{\mathsf 0}\,.
\end{alignat}
Equations~(\ref{holonomic}) in fact merely ensure that only holonomic enter into consideration:
\begin{equation*}
\frac{d\sX}{dt} =\sV\,, \quad\frac{d\sV}{dt}= \sV'\,.
\end{equation*}
Equations~(\ref{OstrohradDefinition}) give the definition of Ostrohrad\softm skyj momenta:\footnote{Although we use bold character to denote the two-component array~$\sP$, we have to  warn the Reader that, whereas $\sR$ transforms like a vector, $\sP$ does not, quite the same way as neither does~$\sX$.}
\begin{equation}\label{matsyuk:Legendre-Ostrohradskyj}
    \sP=\frac{\partial L}{\partial \sV}-\frac{d\sR}{dt}\,,\quad \sR=\frac{\partial L}{\partial \sV'}\,.
\end{equation}
Equation~(\ref{Euler-Poisson}) is the variational equation of motion (the Euler-Poisson vector equation)
\begin{equation*}
    \frac{\partial L}{\partial \sX}-\frac{d}{dt}\left(\frac{\partial L}{\partial \sV}-\frac{d}{dt}\frac{\partial L}{\partial \sV'}\right)=\boldsymbol{\mathsf 0}\,.
\end{equation*}

As soon as any Lagrange function that corresponds to the equation~(\ref{10}), should be of affine type with respect to the accelerations, the definition~(\ref{matsyuk:HamFunction}) reduces to
\begin{equation}\label{matsyuk:HamReduced}
    H=\sP.\sV-\mu \sqrt{1+\sv_i\sv^i} \,.
\end{equation}

Let us take the Lagrange function~(\ref{L1}) in place of~$L$.
 Then equation~(\ref{OstrohradDefinition}) in its first component produces
\begin{equation}\label{matsyuk:r1=0}
   \frac{d\sr_1}{dt} = -\frac{\partial H}{\partial \sv^1}\,,\quad  \sr_1=\mathsf 0\,.
\end{equation}
Now the consistency condition for the pair of equations~(\ref{matsyuk:r1=0}) reads
\begin{equation}\label{matsyuk:r1consistency}
    \frac{\partial H}{\partial \sv^1}=\mathsf 0\,,
\end{equation}
which, together with
\begin{equation}\label{matsyuk:dH/dv'^2}
    \frac{\partial H}{\partial \sv'^2}=\mathsf 0
\end{equation}
from~(\ref{OstrohradDefinition}),
 allows us to get rid not only of the variable $\sv'^2$ ($\sv'^1$ is already out of game by $\dfrac{\partial L_1}{\partial \sv'^1}$=0), but also of $\sv^1$, thus reducing the overall number of canonical variables by two (i.e. ignoring the conjugated pair $(\sv^1, \sr_1)$).
The theory is non-degenerate in the sense that the Hessian
{
\setlength{\arraycolsep}{2.3\jot}\renewcommand{\arraystretch}{1.5}
\setlength{\mathindent}{0pt}
\setlength{\multlinegap}{0pt}
\begin{multline*}
    \begin{Vmatrix}&
      \frac{\partial^2 L_1}{\partial \strut{\sv'}^2\partial {\sv'}^2} & \frac{\partial^2 L_1}{\partial \strut {\sv'}^2\partial {\sv^1}} &\\
      &
      \frac{\partial^2 L_1}{\partial \strut\sv^1\partial {\sv'}^2} & \frac{\partial^2 L_1}{\partial \strut \sv^1\partial {\sv^1}} &
    \end{Vmatrix}\\
    =
    \begin{Vmatrix}
      0 & -\frac{1}{\strut(1 + -\sv_i\sv^i)^{3/2}} \\
      \;-\frac{1}{\strut(1+\sv_i\sv^i)^{3/2}} & 3\frac{\strut{\sv'}^2\sv_1}{\strut(1+\sv_i\sv^i)^{5/2}}+\mu\eta_1\frac{\strut1+\sv_2\sv^2}{\strut(1+\sv_i\sv^i)^{3/2}}
      \;
    \end{Vmatrix}
\end{multline*}
}
is of the rank~$2$.

Let us solve equation~(\ref{matsyuk:dH/dv'^2}) for~$\sv^1$. With $L_1$ in place of~$L$, equation~(\ref{matsyuk:dH/dv'^2}) reads
\begin{equation}\label{matsyuk:solvev1}
    \sr_2=-\frac{\sv^1}{\sqrt{1+\sV\bcdot\sV}\left(1+\sv_2\sv^2\right)}\,.
\end{equation}
Now take the square of both sides:
\begin{equation*}
    \eta_1\sr_2\sr_2\left(1+\sV\bcdot\sV\right)\left(1+\sv_2\sv^2\right)^2=\sv_1\sv^1\,,
\end{equation*}
from where by collecting like terms,
\begin{multline}\label{matsyuk:v1v1}
   \sv_1\sv^1\left(1-\eta_1\sr_2\sr_2\Bigl(1+\sv_2\sv^2\Bigr)^2\right)\\
   =\eta_1\sr_2\sr_2\Bigl(1+\sv_2\sv^2\Bigr)^3.
   \end{multline}
Let us construct the expression $(1+\sV\bcdot\sV)$ from~(\ref{matsyuk:v1v1}) as follows:
\begin{alignat}{3}
    &(1+\sV\bcdot\sV)\mspace{-4mu}&\biggl(&1&-&\eta_1\sr_2\sr_2\Bigl(1+\sv_2\sv^2\Bigr)^2\biggr)
\notag\\
    &&=&1&-&\eta_1\sr_2\sr_2\Bigl(1+\sv_2\sv^2\Bigr)^2
\notag\\
    &&&&+&\eta_1\sr_2\sr_2\Bigl(1+\sv_2\sv^2\Bigr)^3+\sv_2\sv^2
\notag\\
    &&&&-&\eta_1\sv_2\sv^2\sr_2\sr_2\Bigl(1+\sv_2\sv^2\Bigr)^2
\notag\\
    &&=&1&+&\sv_2\sv^2+\Bigl(1+\sv_2\sv^2\Bigr)^2\cdot0
\notag\\\label{matsyuk:1+vv}
    &&=&1&+&\sv_2\sv^2\,.
\end{alignat}
The expression for the variable~$\sv'^2$ might have been obtained from~(\ref{matsyuk:r1consistency}), but actually we are not interested in it as far as we are going to use directly  formula~(\ref{matsyuk:HamReduced}) rather than~(\ref{matsyuk:HamFunction}).
By~(\ref{matsyuk:solvev1}) and~(\ref{matsyuk:1+vv}) the Hamiltonian~(\ref{matsyuk:HamReduced}) becomes
\begin{multline}\label{matsyuk:H1}
    H=\sp_2\sv^2-\sp_1\sr_2\frac{\Bigl(1+\sv_2\sv^2\Bigr)^{3/2}}{\sqrt{1-\eta_1\sr_2\sr_2\Bigl(1+\sv_2\sv^2\Bigr)^2}}
    \\[1\jot]
    -\mu\sqrt{\frac{1+\sv_2\sv^2}{1-\eta_1\sr_2\sr_2\Bigl(1+\sv_2\sv^2\Bigr)^2}}\,.
\end{multline}

The Poisson structure is implemented by the Poisson bracket
\[
\Big\{F,G\Big\}_{\sP,\sR} = \frac{\partial F}{\partial \x^i} \frac{\partial G}{\partial
\sfp_i} - \frac{\partial F}{\partial \sp_i} \frac{\partial G}{\partial \x^i} +
\frac{\partial F}{\partial \sv^2} \frac{\partial G}{\partial \sr_2} -
\frac{\partial F}{\partial \sr_2} \frac{\partial G}{\partial \sv^2},
\]
and the generalized Hamilton equations read:

\parbox{3cm}{
\begin{eqnarray*}
\frac{d\x^i}{dt}   & =  &   \Big\{\x^i,H\Big\}_{\sP,\sR}
\\  \frac{d\sv^2}{dt} & =  &    \Big\{\sv^2,H\Big\}_{\sP,\sR}
\end{eqnarray*}
}
\quad
\parbox{3cm}{
\begin{eqnarray*}
\frac{d\sp_i}{dt}   & =  &   \Big\{\sp_i,H\Big\}_{\sP,\sR}
\\ \frac{d\sr_2}{dt} & =  &   \Big\{\sr_2,H\Big\}_{\sP,\sR}\,.
\end{eqnarray*}
}

If we had started with~(\ref{L2}), then definition~(\ref{matsyuk:HamFunction}) would have changed to new variables, $\tsP$, $\tsR$, and a new function~$\tH$,
\begin{equation}\label{matsyuk:HamFunctionTilde}
    \tH=\tsP.\sV+\tsR.\sV'-L_2\,.
\end{equation}
In place of (\ref{matsyuk:r1=0}), (\ref{matsyuk:r1consistency}), and~(\ref{matsyuk:dH/dv'^2}) we should have had
\begin{equation}\label{matsyuk:r2=0}
   \frac{d\tsr_2}{dt} = -\frac{\partial \tH}{\partial \sv^2}\,,\quad  \tsr_2=\mathsf 0\,,
\end{equation}
\begin{equation*}
    \frac{\partial \tH}{\partial \sv^2}=\mathsf 0\,.
\end{equation*}
and
\begin{equation}\label{matsyuk:dH/dv'1}
    \frac{\partial H}{\partial \sv'^1}=\mathsf 0
\end{equation}
from~(\ref{OstrohradDefinition}).
Formul{\ae} (\ref{matsyuk:solvev1}) and~(\ref{matsyuk:1+vv}) would have been substituted by
\begin{equation}\label{matsyuk:solvev2}
    \tsr_1=\frac{\sv^2}{\sqrt{1+\sV\bcdot\sV}\left(1+\sv_1\sv^1\right)}
\end{equation}
and
\begin{equation}\label{matsyuk:1+vvTilde}
    (1+\sV\bcdot\sV)\biggl(1-\eta_2\tsr_1\tsr_1\Bigl(1+\sv_1\sv^1\Bigr)^2\biggr) =1+\sv_1\sv^1\,.
\end{equation}
The theory becomes non-degenerate with the Hessian
{
\setlength{\arraycolsep}{2.3\jot}\renewcommand{\arraystretch}{1.5}
\setlength{\mathindent}{0pt}
\setlength{\multlinegap}{0pt}
\begin{multline*}\setlength{\multlinegap}{0pt}
    \begin{Vmatrix}&
      \frac{\partial^2 L_2}{\partial \strut{\sv'}^1\partial {\sv'}^1} & \frac{\partial^2 L_2}{\partial \strut {\sv'}^1\partial {\sv^2}} &\\
      &
      \frac{\partial^2 L_2}{\partial \strut\sv^2\partial {\sv'}^1} & \frac{\partial^2 L_2}{\partial \strut \sv^2\partial {\sv^2}} &
    \end{Vmatrix}\\
    =
    \begin{Vmatrix}
      0 & \frac{1}{\strut(1 + -\sv_i\sv^i)^{3/2}} \\
      \;\frac{1}{\strut(1+\sv_i\sv^i)^{3/2}} & -3\frac{\strut{\sv'}^1\sv_2}{\strut(1+\sv_i\sv^i)^{5/2}}+\mu\eta_2\frac{\strut1+\sv_1\sv^1}{\strut(1+\sv_i\sv^i)^{3/2}}
      \;
    \end{Vmatrix}
\end{multline*}
}
With (\ref{matsyuk:solvev2}) and~(\ref{matsyuk:1+vvTilde}) the new Hamilton function~(\ref{matsyuk:HamFunctionTilde}) reads
\begin{multline}\label{matsyuk:H2}
    \tH=\tsp_1\sv^1+\tsp_2\tsr_1\frac{\Bigl(1+\sv_1\sv^1\Bigr)^{3/2}}{\sqrt{1-\eta_2\tsr_1\tsr_1\Bigl(1+\sv_1\sv^1\Bigr)^2}}
    \\[1\jot]
    -\mu\sqrt{\frac{1+\sv_1\sv^1}{1-\eta_2\tsr_1\tsr_1\Bigl(1+\sv_1\sv^1\Bigr)^2}}\,.
\end{multline}
The new Poisson structure would have been given by
\begin{equation}\label{matsyuk:PoissonTilde}
    \Big\{F,G\Big\}_{\tsP,\tsR} = \frac{\partial F}{\partial \x^i} \frac{\partial G}{\partial
\tsp_i} - \frac{\partial F}{\partial \tsp_i} \frac{\partial G}{\partial \x^i} +
\frac{\partial F}{\partial \sv^1} \frac{\partial G}{\partial \tsr_1} -
\frac{\partial F}{\partial \tsr_1} \frac{\partial G}{\partial \sv^1},
\end{equation}
and the new generalized Hamilton equations would have read:

\parbox{3cm}{
\begin{eqnarray*}
\frac{d\x^i}{dt}   & =  &   \Big\{\x^i,H\Big\}_{\tsP,\tsR}
\\  \frac{d\sv^1}{dt} & =  &    \Big\{\sv^1,H\Big\}_{\tsP,\tsR}
\end{eqnarray*}
}
\quad
\parbox{3cm}{
\begin{eqnarray*}
\frac{d\sp_i}{dt}   & =  &   \Big\{\sp_i,H\Big\}_{\tsP,\tsR}
\\ \frac{d\sr_1}{dt} & =  &   \Big\{\sr_1,H\Big\}_{\tsP,\tsR}\,.
\end{eqnarray*}
}

The Hamilton function in~(\ref{matsyuk:HamFunction}) is defined on the space $T^2Q\times_{TQ}T^*\big(TQ\big)$ with coordinates $\sX$, $\sV$, $\sV'$, $\sP$, $\sR$ and may be lifted to the so-called \textit{unified phase space}~\cite{Prieto-Martinez} $T^3Q\times_{TQ}T^*\big(TQ\big)$ with coordinates $\sX$, $\sV$, $\sV'$, $\sV''$, $\sP$, $\sR$ by the projection ignoring~$\sV''$. The \textit{Legendre--Ostrohrad\softm skyj map} $T^3Q\to T^*\big(TQ\big)$ over~$TQ$ (see again~\cite{Prieto-Martinez}), defined by formul{\ae}~(\ref{matsyuk:Legendre-Ostrohradskyj}), in our case of affine Lagrangian actually is defined on the space~$T^2Q$ and may be thought of as a graph in the space $T^2Q\times_{TQ}T^*\big(TQ\big)$. It may be described by a pair of applications, $\sP=\pi(\sV,\sV')$, and~$\sR=\rho(\sV,\sV')$. Given another Hamilton function,~$\tH$, one arrives at another Legendre--Ostrohrad\softm skyj map,~$\tsP=\tpi(\sV,\sV')$, and~$\tsR=\trho(\sV,\sV')$.
As far as~(\ref{matsyuk:F}) holds, and assuming $\dfrac{\partial F}{\partial \sV'}=0$, from~(\ref{matsyuk:Legendre-Ostrohradskyj}) one gets
\begin{align}
    \tsR&\stackrel{\mathrm{def}}{=}\frac{\partial L_2}{\partial \sV'}\equiv\frac{\partial L_1}{\partial \sV'}+\frac{\partial}{\partial \sV'}\left(\sV.\frac{\partial F}{\partial \sX}+\sV'.\frac{\partial F }{\partial \sV}\right)
    \notag\\\label{matsyuk:r+phi}
    &=\sR+\boldsymbol\phi,
    \\\label{matsyuk:dF/dv'}
    &\quad\mbox{where }\boldsymbol\phi\stackrel{\mathrm{def}}{=}\frac{\partial F}{\partial \sV}\,,
    \\
    \tsP&\stackrel{\mathrm{def}}{=}\frac{\partial L_2}{\partial \sV}-\frac{d\tsR}{dt}=\sP+\left(\frac{\partial }{\partial \sV}-\frac{d}{dt}\frac{\partial }{\partial \sV'}\right)\frac{dF}{dt}
    \notag\\
    &=\sP+\frac{\partial F}{\partial \sX}\,.
    \notag
\end{align}
In our example $\dfrac{\partial F}{\partial \sX}=0$, so one can drop the diacritic \text{tilde} over $\sP$ and~$\pi$. From~(\ref{matsyuk:r+phi}) one can obtain the expression for $\phi_1$, using~(\ref{matsyuk:solvev2}) together with the second equation of~(\ref{matsyuk:r1=0}), and the expression for~$\phi_2$, using~(\ref{matsyuk:solvev1}) accompanied by the second equation of~(\ref{matsyuk:r2=0})):
\begin{subequations}
\label{phi_both}
\renewcommand{\theequation}{\theparentequation.\ms\arabic{equation}}
\begin{align}\label{phi1}
    \phi_1&=\tsr_1=\frac{\sv^2}{\sqrt{1+\sV\bcdot\sV}\bigl(1+\sv_1\sv^1\bigr)}\,,
    \\\label{phi2}
    \phi_2&=-\sr_2=\frac{\sv^1}{\sqrt{1+\sV\bcdot\sV}\bigl(1+\sv_2\sv^2\bigr)}\,.
\end{align}
\end{subequations}
One can easily check
\begin{equation}\label{matsyuk:phi_i,j}
    \frac{\partial \phi_1}{\partial \sv^2}=\frac{1}{(1+\sV\bcdot\sV)^{3/2}}=\frac{\partial \phi_2}{\partial \sv^1}\,,
\end{equation}
which should be obvious from~(\ref{matsyuk:dF/dv'}).

Let us consider the  following diagram
\begin{equation}\label{matsyuk:diagram}
\hfil
\xy
\xymatrix{
&\{\sV,\sV'\}\ar[dl]_{(id,\pi,\rho)}\ar[dr]^{(id,\pi,{\trho})}&
\\
\{\sV,\sP,\sR\}\ar[dr]_H\ar[rr]^{(id,id,\Phi)}&&\{\sV,\sP,{\tsR}\}\ar[dl]^{\tH}
\\
&\R&
}
\endxy
\end{equation}
where the map $\Phi$ is given by $\Phi(\sV,\sP,\sR)=\sR+\boldsymbol{\phi}(\sV)$.
\begin{prop}
The  diagram~(\ref{matsyuk:diagram}) commutes
\end{prop}
\noindent\textit{Proof.\/}
By the second formula in~(\ref{matsyuk:r1=0}), together with~(\ref{phi2}),
in the upper triangle to the left we get
\begin{alignat}{3}
    \rho_1(\sV,\sV')&=0;&\quad& \rho_2(\sV,\sV')&&=-\phi_2\,,\notag
\\\intertext{whereas by~(\ref{phi1}) and~(\ref{matsyuk:r2=0}) to the right we have}
\trho_1(\sV,\sV')&=\phi_1;&&\trho_2(\sV,\sV')&&=0\,.\label{matsyuk:upper right diagram arrow}
\end{alignat}
On the right by~(\ref{phi1}) one calculates
{
\setlength{\mathindent}{0pt}
\begin{align*}
    \Bigl(\Phi\circ\rho\Bigr)_1(\sV,\sV')&=\Phi_1\Bigl(\rho_1(\sV),\rho_2(\sV)\Bigr)=0+\phi_1=\phi_1\,,
    \\
    \Bigl(\Phi\circ\rho\Bigr)_2(\sV,\sV')&=\Phi_2\Bigl(\rho_1(\sV),\rho_2(\sV)\Bigr)=-\phi_2+\phi_2=0\,,
\end{align*}
which coincides with~(\ref{matsyuk:upper right diagram arrow}).
}

In the lower triangle we compute the composition
\begin{multline*}
\tH\circ(id,id,\Phi)(\sV,\sP,\sR)=\sP.\sV+\Phi(\sV,\sP,\sR).\sV'-L_2
\\
=\sP.\sV+\sR.\sV'+\boldsymbol\phi.\sV'-L_2=\sP.\sV+\sR.\sV'-L_1
\\
=H(\sV,\sP,\sR)\,,
\end{multline*}
because $\boldsymbol\phi.\sV'=L_2-L_1$ in view of~(\ref{phi_both}).

\rightline{Q.E.D.}

Let us turn to the notion of canonical transformations. Consider some space of conjugated pairs of variables $\{Q^A, P_A\}$ with a Poisson bracket
\begin{equation*}
\{F,G\}_{P,Q}=\frac{\partial F}{\partial Q^A} \frac{\partial G}{\partial
P_A} - \frac{\partial F}{\partial P_A} \frac{\partial G}{\partial Q^A}
\end{equation*}
and a transformation
\begin{equation}\label{matsyuk:canonical}
\bigl(Q^A, P_A\bigr)\mapsto\bigl(\tQ^{\strut A}, \tP_A\bigr).
\end{equation}
\begin{define}\label{definition}
Transformation~(\ref{matsyuk:canonical}) is called \textit{canonical} if the following two equivalent conditions hold~\cite{Leon,Gitman}:
\begin{list}{\textbf{\arabic{listcounter}}.}{\usecounter{listcounter}
\setlength{\leftmargin}{0em}
\setlength{\itemindent}{\parindent}
}
\item\label{matsyuk:list1}
The Jacobian matrix of the inverse transformation to~(\ref{matsyuk:canonical}) reads
{\setlength{\arraycolsep}{2.3\jot}\renewcommand{\arraystretch}{1.5}
\begin{equation}\label{matsyuk:inverse_jacobi}
    \begin{Vmatrix}
      \left(\dfrac{\partial \tbP}{\partial \bP}\right)^T & -\left(\dfrac{\partial \tbQ}{\partial \bP}\right)^T \\
      -\left(\dfrac{\partial \tbP}{\partial \bQ}\right)^T & \left(\dfrac{\partial \tbQ}{\partial \bQ}\right)^T
    \end{Vmatrix}
\end{equation}
}
where superscript~$T$ denotes the transposition of the embraced matrix.
\item
\label{matsyuk:list2}The Poisson bracket of the new conjugate pairs calculated with respect to the old variables does not change:
\begin{equation}\label{matsyuk:canonical_Gitman}
\begin{gathered}
    \{\tQ^A,\tQ^B\}_{P,Q}=0\,;\; \{\tP_A,\tP_B\}_{P,Q}=0\,;
    \\[2\jot]
    \{\tQ^A,\tP_B\}_{P,Q}=\delta^A_B\,.
\end{gathered}
\end{equation}
\end{list}
\end{define}
\begin{prop}
Let $\eta:\{\sX,\sV,\sP,\sR\}\to\{\sX,\sV,\sP,{\tsR}\}$ be defined as $\eta=(id,id,id,\Phi)$ with $\Phi$ from the diagram~(\ref{matsyuk:diagram}). Then $\eta$ is canonical.
\end{prop}

\noindent\textit{Proof.\/} We shall prove both properties in Definition~(\ref{definition}).
\begin{list}{}{
\setlength{\leftmargin}{0em}
\setlength{\itemindent}{\labelindent}
}
\item[\textbf{Proof of property}~(\ref{matsyuk:list1}).]
The Jacobian matrix of~$\eta$ is
{\setlength{\arraycolsep}{2.3\jot}\renewcommand{\arraystretch}{1.5}
\begin{equation}\label{matsyuk:jacobi_eta}
\begin{Vmatrix}
1&0&0&0&0&0&0&0
\\
0&1&0&0&0&0&0&0
\\
0&0&1&0&0&0&0&0
\\
0&0&0&1&0&0&0&0
\\
0&0&0&0&1&0&0&0
\\
0&0&0&0&0&1&0&0
\\
0&0&
\dfrac{\partial \phi_1}{\partial \sv^1}&\dfrac{\partial \phi_1}{\partial \sv^2}
&0&0&1&0
\\
0&0&
\dfrac{{}^{\strut}\partial \phi_2}{\partial \sv^1}&\dfrac{\partial \phi_2}{\partial \sv^1}
&0&0&0&1
\end{Vmatrix}
    \end{equation}

\noindent Obviously its determinant is~$1$.

The inverse to~(\ref{matsyuk:jacobi_eta}) is
\begin{equation}\label{matsyuk:inverse_jacobi_eta}
\begin{Vmatrix}
1&0&0&0&0&0&0&0
\\
0&1&0&0&0&0&0&0
\\
0&0&1&0&0&0&0&0
\\
0&0&0&1&0&0&0&0
\\
0&0&0&0&1&0&0&0
\\
0&0&0&0&0&1&0&0
\\
0&0&
-\dfrac{\partial \phi_1}{\partial \sv^1}&-\dfrac{\partial \phi_1}{\partial \sv^2}
&0&0&1&0
\\
0&0&
-\dfrac{{}^{\strut}\partial \phi_2}{\partial \sv^1}&-\dfrac{\partial \phi_2}{\partial \sv^1}
&0&0&0&1
\end{Vmatrix}
\end{equation}
}

In order to apply~(\ref{matsyuk:inverse_jacobi}) to~(\ref{matsyuk:inverse_jacobi_eta}) we put $Q^A=(\x^i,\sv^i)$ and $P_A=(\sp_i,\sr_i)$ and then we profit from~(\ref{matsyuk:phi_i,j}) to check that~(\ref{matsyuk:inverse_jacobi}) holds.

\rightline{Q.E.D.}
\item[\textbf{Proof of property}~(\ref{matsyuk:list2}).]
By the definition of~$\eta$ we are led to put $\tQ^A=Q^A$, $\tsp_i=\sp_i$, and $\tsr_i=\sr_i+\phi_i$, so that
{
\setlength{\mathindent}{0pt}
\begin{gather*}
    \{\tQ^A,\tQ^B\}=\{Q^A,Q^B\}=0\,;
    \\
   \{\tsv^i,\tsp_j\}=\{\sv^i,\sp_j\}=0\,;\;\{\tsp_i,\tsp_j\}=\{\sp_i,\sp_j\}=0\,;
    \\
  \{\tsx^i,\tsr_j\}=\{\x^i,\sr_j+\phi_j\}=\{\x^i,\phi_j\}=0\,;
  \\
  \{\tsp_i,\tsr_j\}=\{\sp_i,\sr_j+\phi_j\}
       =\{\sp_i,\phi_j\}=0\,;
    \\
\{\tsx^i,\tsp_j\}=\{\x^i,\sp_j\}=\delta^i_j\,;
\\
\{\tsv^i,\tsr_j\}=\{\sv^i,\sr_j+\phi_j\}=\{\sv^i,\sr_j\}=\delta^i_j
    \\
\begin{split}
   \{\tsr_i,\tsr_j\}&=\{\sr_i+\phi_i,\sr_j+\phi_j\}=\{\sr_i,\phi_j\}+\{\phi_i,\sr_j,\}
   \\
   &=-\frac{\partial \phi_j}{\partial \sv^k}\delta^k_i+\frac{\partial \phi_i}{\partial \sv^k}\delta^k_j=0
   \\
   &
\qquad\qquad\qquad\qquad\text{on the strength of~(\ref{matsyuk:phi_i,j})}\,.
\end{split}
\end{gather*}
}
This agrees with~(\ref{matsyuk:canonical_Gitman}).

\rightline{Q.E.D.}
\end{list}

\end{document}